# Building an AI-ready RSE Workforce

Ying Zhang[1], Matthew A. Gitzendanner[1], Dan S. Maxwell[1], Justin W. Richardson[1], Kaleb E. Smith[2], Eric A. Stubbs[1], Brian J. Stucky[1], Jingchao Zhang[2], Erik Deumens[1]

1: University of Florida, {yingz, magitz, danielmaxwell, jstrds, ericeric, stuckyb, deumens}@ufl.edu

2: Nvidia, {kasmith, jingchaoz}@nvidia.com

**ABSTRACT**

Artificial Intelligence has been transforming industries and academic research across the globe, and research software development is no exception. Machine learning and deep learning are being applied in every aspect of the research software development lifecycle, from new algorithm design paradigms to software development processes. In this paper, we discuss our views on today's challenges and opportunities that AI has presented on research software development and engineers, and the approaches we, at the University of Florida, are taking to prepare our workforce for the new era of AI.

## 1 Introduction

In the past decade, Artificial Intelligence (AI) has become ubiquitous in everyday life, from smart gadgets and personalized online shopping to autonomous driving. AI embodies intelligence in machines, automates mundane tasks, and increases productivity. As the driving force behind many recent technological innovations, AI has attracted great interest in academic communities and transformed research approaches across disciplines. Research software is at the core of these technology innovations and academic research. Hence, it is paramount for Research Software Engineers (RSEs) to adopt emerging AI-oriented approaches in algorithm development and data analytics and to transform software development paradigm by leveraging AI tools to streamline every aspect of the software development cycle. Here, we discuss new initiatives at the University of Florida (UF) that we hope will help RSEs more easily build these skills.

## 2 AI Impact on research and research software

The applications of AI in research have grown to include generating hypotheses, data analytics, and conducting experiments in many research disciplines. Machine learning (ML), and deep learning (DL) in particular, are being widely applied in many research fields, such as drug discovery [1], quantum chemistry [2], and communications and journalism [3]. AI-enabled software platforms, tools, and libraries are being developed and incorporated into research software packages and workflows in areas such as genomics research, precision medicine, natural language processing, human-computer interaction, and data ethics. Elsevier's 2019 Research Futures report [4] discusses how AI is shaping the research landscape in these significant aspects:

- Data science, particularly AI, shapes data access, collection, and information extraction at unprecedented scale and speed. Developments in AI also enable data-driven hypotheses and research approaches. Despite reduced public funding, the volume of research has been increasing steadily.
- AI-powered technologies provide new methods of generating and communicating results and supporting peer review procedures.

With a campus-wide AI initiative that began in the spring of 2020, UF is making AI the centerpiece of a major long-term effort to become a national leader in AI research, education, and workforce development. To support this bold vision, UF has committed to hiring 100 new AI faculty across disciplines and deployed a new $70 million supercomputer, HiPerGator AI, currently ranked at No. 22 on the TOP500 list [5]. As an incentive to researchers, UF launched the AI Catalyst Fund in 2020 to jump-start interdisciplinary research and collaboration in AI. More than 130 proposals from across campus were received, and 20 faculty teams from 11 UF colleges were awarded the grant. To facilitate research and collaboration, UF

implemented business plans to provide access to HiPerGator AI to the UF community and all universities in the state of Florida. This AI initiative presented an incredible opportunity for researchers and software developers across the state to take advantage of the world-class AI supercomputer for innovative research and development. Beyond UF, academic hiring centered on AI/ML/Data Mining has also steadily risen worldwide [6].

## 3  AI Impact on Research Software Development

AI technology is transforming software development and improving its productivity, quality, and speed. AI and ML are being incorporated into software development cycles [7] in three major phases:

1) **Project planning**: Software requirement gathering, project time and budgeting, and task planning require large amounts of human intervention. AI/ML methods can be utilized in data-driven user requirement analysis, tool selections, task assignment, and cost optimization.
2) **Software design and development**: Software design often requires specialized knowledge and skills. AI Design Assistant (AIDA) tools can automate complex procedures and adapt solutions according to user needs in an agile fashion. Deep learning strategies such as natural language processing and neural networks have also been used for auto-coding [8].
3) **Software testing and integration**: Testing, maintenance, and integration are essential steps in software development. It can also be extremely repetitive and time-consuming. AI tools can be employed for regression testing, software update, and integrations, significantly reducing staff costs.

To incorporate AI in software development circles, RSEs must build up fundamental knowledge about AI, as well as develop a different mindset to adjust to the paradigm.

## 4  Empowering RSEs with AI Skills

To keep up with the rapid application of AI in research and software development, RSEs must adapt their knowledge base and skill sets to meet future challenges and stay competitive. To bridge the gap, we need to ask what types of skills RSEs must acquire and how to develop the RSE workforce for the future. In Software 2.0 [9], Dr. Andrej Karpathy laid out his grand vision for software development in the age of AI. The key components are problem definition, data collection and curation, and model learning and management. A detailed discussion of various AI skills required for RSEs is beyond the scope of this paper. We focus instead on new education initiatives at UF that we believe will be well-suited for RSEs and related professionals to develop cutting-edge AI skills.

The UF AI initiative includes a public-private partnership with Nvidia and the establishment of the United States' first Nvidia AI Technology Center (NVAITC), located at UF. This partnership expands UF's research strength in AI and provides a unique opportunity for AI training and workforce development. At the college level, UF is integrating AI across the curriculum with modules for specific technical and research-focused domains. Nvidia's Deep Learning Institute (DLI) offers a wide range of AI teaching and training resources. These include AI/ML/DL fundamentals, GPU accelerated computing and data science, and research domain-focused applications such as medical imaging, robotics, and deep learning at scale.

In the past year, partnering with Nvidia, UF Research Computing offered eight DLI workshops with real-world AI use cases, nine hands-on short courses on specific AI fundamental topics, and ten informal Birds-of-a-Feather sessions on practical aspects of AI applications. We are also developing an online AI education program that offers students certificates in various AI skills from basic to advanced levels. These training and education opportunities are tailored to RSE's skill sets and provide a road map to develop AI competence at their own pace and levels.

## 5  Conclusion

Without any doubt, AI will be a substantial driving force behind science and research in the years to come and reshape software development. To stay productive and competitive, RSEs must acquire AI knowledge and skills to be able to navigate the AI landscape and rise to the challenges and opportunities of the future.